# Computing Best-Response Strategies
# in Infinite Games of Incomplete Information*

**Daniel M. Reeves**    **Michael P. Wellman**
University of Michigan Artificial Intelligence Lab.
1101 Beal Avenue, Ann Arbor, MI 48109-2110 USA
http://ai.eecs.umich.edu/people/{dreeves,wellman}/
{dreeves,wellman}@umich.edu

## Abstract

We describe an algorithm for computing best-response strategies in a class of two-player infinite games of incomplete information, defined by payoffs piecewise linear in agents' types and actions, conditional on linear comparisons of agents' actions. We show that this class includes many well-known games including a variety of auctions and a novel allocation game. In some cases, the best-response algorithm can be iterated to compute Bayes-Nash equilibria. We demonstrate the efficacy of our approach on existing and new games.

## 1 Introduction

Game theory was founded by von Neumann and Morgenstern (1947) to study situations in which multiple agents (players) interact in order to each maximize an objective (payoff) function determined not only by their own actions but also the actions of other players. In a game of incomplete information, the payoffs also depend on information that is private to the individual agents. This information is known as an agent's *type*. We consider one-shot games—games in which an agent chooses a single action based only on knowledge of its own type, the payoff function, and the distribution from which types are drawn. We assume that the available actions, payoff function, and type distributions are common knowledge.[1] A strategy in this context is a mapping from the set of types to the set of actions. The case where these sets are finite (and especially when there are no types—complete information) has been well studied in the computational game theory literature. In Sections 2 and 6 we discuss existing algorithms and software tools for solving finite games. But for incomplete information games with a continuum of actions available to the agents—known as infinite games—we know of no available algorithms, though many particular infinite games of incomplete information have been solved in the literature.

We have so far left unsaid what we mean by solving a game. For the case of a single agent, a solution is a policy that maximizes the agent's expected utility (payoff). In the case of multiple agents, Nash (1951) proposed a solution concept now known as the *Nash equilibrium*, and proved that for finite games (as long as agents can play mixed strategies—i.e., randomize among actions) such an equilibrium always exists. A Nash equilibrium is a profile of strategies such that each strategy is a best response to the rest of the profile. That is, each agent maximizes its own utility given the strategies played by the others.[2]

This definition of Nash equilibrium invites an obvious algorithm for finding one: start with a profile of seed strategies and iteratively compute best-response profiles until a fixed-point is reached. This process (when it converges) yields a profile which is a best response to itself and thus a Nash equilibrium.

The notion of Nash equilibrium needs to be generalized slightly for the case of incomplete information games. Another seminal game theorist, Harsanyi (1967), introduced the concept of agent types and used it to define a *Bayesian game*. A Bayesian game is specified by a set of types $T$, a set of strategies $S$, a probability distribution $F$ over types, and a payoff function $P$. Harsanyi defines a *Bayes-Nash equilibrium* (sometimes known as a Bayesian equilibrium) as the simple Nash equilibrium of the non-Bayesian game with set of strategies being the set of functions from $T$ to $S$, and the payoff function being the expectation of $P$ with respect to $F$.

---

[1] A fact is common knowledge (Fagin et al., 1995) if everyone knows it, everyone knows that everyone knows it, ad infinitum.

[2] The limitations of Nash equilibrium as a solution concept have been well studied in the literature, particularly the problem of what agents will do in the face of multiple equilibria (van Damme, 1983) yet finding Nash equilibria remains fundamental to the analysis of games.

## 2 Finite Game Approximations

GAMBIT (McKelvey et al., 1992) is a software package incorporating several algorithms for solving finite games (McKelvey and McLennan, 1996). The original 1992 implementation of GAMBIT was limited to normal form games.[3] GALA (Koller and Pfeffer, 1997) introduced constructs for schematic specification of extensive-form games, with a solver exploiting recent algorithmic advances (Koller et al., 1996). GAMBIT subsequently incorporated these GALA features, and currently stands as the standard in general finite game solvers.

We employ a standard first-price sealed-bid auction (FPSB) to compare our approach for the full infinite game to a discretized version amenable to finite game solvers. Consider a discretization of (FPSB) with two players, nine types, and nine actions. Both players have a randomly (uniform) determined valuation ($t$) from the set $\{0,\ldots,8\}$ and the available actions ($a$) are to bid an integer in the set $\{0,\ldots,8\}$. Let $[p,a;a']$ denote the mixed strategy of performing action $a$ with probability $p$, and $a'$ with probability $1-p$. (There happened never to be mixed strategies with more than two actions.) GAMBIT solves this game exactly, finding the following Bayes-Nash equilibrium:[4]

| $t$: | 0 | 1 | 2 | 3 | 4 | 5 | 6 | 7 | 8 |
|---|---|---|---|---|---|---|---|---|---|
| $a(t)$: | 0 | 0 | 1 | 1 | 2 | [.455,2;3] | 3 | 3 | [.727,3;4] |

This result is indeed close to the unique Bayes-Nash equilibrium, $a(t) = t/2$, of the corresponding infinite game (see Section 5.1). However, it varies asymmetrically, which implies that the discretized version represents a qualitatively different game.

We observe a similar asymmetric divergence in another discrete approximation. Consider the FPSB over a continuous type range, distributed uniformly, but with actions restricted to the form $a(t) = wt$, with $w$ a choice parameter. (This is a very conservative comparison case in that it eliminates all but a class of strategies that includes the known equilibrium.) For the continuous action range $w \in [0,1]$, $w = 1/2$ constitutes a unique equilibrium. When $w$ is limited to discrete values on a grid, however, additional equilibria emerge. For example, when only increments of 0.05 are allowed, both $w = 0.45$ and $w = 0.50$ represent symmetric equilibrium policies.

The point of these examples is that whereas solving finite approximations to an infinite game can be instructive, it can also produce misleading results. In Section 5.1 we show how our method immediately finds the solution to the full infinite FPSB game.

## 3 Infinite Games and Bayes-Nash Equilibria

We consider a class of two-player games, defined by a payoff structure that is analytically restrictive, yet captures many well-known games of interest. Let $t$ denote the subject agent's type and $a$ its action, and $t'$ and $a'$ the type and action of the other agent. We assume that types are scalars drawn from piecewise-uniform probability distributions, and payoff functions take the following form:

$$u(t,a,t',a') = \begin{cases} \theta_1 t + \rho_1 a + \theta'_1 t' + \rho'_1 a' + \phi_1 & \text{if } -\infty < a + \alpha a' < \beta_2 \\ \theta_2 t + \rho_2 a + \theta'_2 t' + \rho'_2 a' + \phi_2 & \text{if } \beta_2 \leq a + \alpha a' \leq \beta_3 \\ \ldots \\ \theta_I t + \rho_I a + \theta'_I t' + \rho'_I a' + \phi_I & \text{if } \beta_I \leq a + \alpha a' \leq +\infty \end{cases} \quad (1)$$

Our class comprises games with payoffs that are linear[5] functions from own type and action, conditional on a linear comparison between own and other agent action. The form is parametrized by $\alpha$, $\beta_i$, $\theta_i$, $\rho_i$, $\theta'_i$, $\rho'_i$, and $\phi_i$, where $i \in \{1,\ldots,I\}$ indexes the comparison case. (We define $\beta_1 \equiv -\infty$ and $\beta_{I+1} \equiv +\infty$ for notational convenience in our algorithm description below.) The $\beta_i, \beta_{i+1}$ regions alternate between open and closed intervals as this is without loss of generality for arbitrary specification of boundary types ('<' vs. '≤') on the regions[6] and in particular allows the implementation of common tie-breaking rules for sealed-bid auctions.

This parameterized payoff function captures many known mechanisms. Table 1 shows the parameter settings for several such games.

Given a game description in this form, we search for Bayes-Nash equilibria through a straightforward iterative process. Starting with a seed strategy profile (typically based on a myopic or naive strategy such as truthful bidding), we repeatedly compute best-response profiles until reaching a fixed-point or cycle. A strategy profile that is a best response to itself is, by definition, a Bayes-Nash equilibrium. We show that this process is effective at finding equilibria for certain games in our class. For all games in our class, the best-response algorithm can be used to verify candidate equilibria or ε-equilibria found by alternate means.

---

[3] *Normal form*, also known as strategic form, lists explicit payoffs for every combination (profile) of agent strategies. In contrast, *extensive form* is a more compact representation for games of imperfect information in which payoffs are given for sequences of actions but only implicitly for combinations of agent strategies (mappings from private information to actions).

[4] The calculation took 90 minutes of cpu time and 17MB of memory on a machine with four 450MHz Pentium 2 processors and 2.5GB RAM, running Linux kernel 2.4.18. When 2 additional types and actions are added to the discretization, a similar equilibrium results, requiring 23 hours of cpu time and 34MB of memory. GAMBIT's algorithm (Koller et al., 1996) is worst-case exponential in the size of the game tree which is itself size $O(n^4)$ in the size of the type/action spaces. Based on this complexity and our timing results, we conclude that we have reached the limit of what GAMBIT's algorithm can compute.

[5] Functions with constant terms are technically *affine* rather than linear but we ignore that distinction from here on.

[6] For example, to specify $u_1$ in $(-\infty,a]$, $u_2$ in $(a,b]$, $u_3$ in $(b,c)$, $u_4$ at $c$, and $u_5$ in $(c,\infty)$, translate to the alternating open/closed specification: $u_1$ in $(-\infty, a-\varepsilon)$, $u_1$ in $[a-\varepsilon, a]$, $u_2$ in $(a,b)$, $u_2$ in $[b,b]$, $u_3$ in $(b,c)$, $u_4$ in $[c,c]$, $u_5$ in $(c,\infty)$.

| Game | $\vec{\theta}$ | $\vec{\rho}$ | $\vec{\theta}'$ | $\vec{\rho}'$ | $\vec{\phi}$ | $\vec{\beta}$ | $\alpha$ |
|---|---|---|---|---|---|---|---|
| FPSB Auction | $0, 1/2, 1$ | $0, -1/2, -1$ | $0, 0, 0$ | $0, 0, 0$ | $0$ | $0, 0$ | $-1$ |
| Vickrey Auction (2nd Price) | $0, 1/2, 1$ | $0, 0, 0$ | $0, 0, 0$ | $0, -1/2, -1$ | $0$ | $0, 0$ | $-1$ |
| Vicious Vickrey Auction | $0, \frac{1-k}{2}, 1-k$ | $k, k/2, 0$ | $-k, -k/2, 0$ | $0, \frac{k-1}{2}, k-1$ | $0$ | $0, 0$ | $-1$ |
| Supply Chain Game | $-1, -1, 0$ | $1, 1, 0$ | $0, 0, 0$ | $0, 0, 0$ | $0$ | $v, v$ | $1$ |
| Bargaining Game  -Seller | $-1, -1, 0$ | $1-k, 1-k, 0$ | $0, 0, 0$ | $k, k, 0$ | $0$ | $0, 0$ | $-1$ |
| -Buyer | $0, 1, 1$ | $0, -k, -k$ | $0, 0, 0$ | $0, 1-k, 1-k$ | $0$ | $0, 0$ | $-1$ |
| All-Pay Auction | $0, 1/2, 1$ | $-1, -1, -1$ | $0, 0, 0$ | $0, 0, 0$ | $0$ | $0, 0$ | $-1$ |
| Voluntary Participation Game | $0, 1$ | $0, -1/2$ | $0, 0$ | $0, 1/2$ | $0, -C/2$ | $C$ | $1$ |
| Shared-Good Auction | $0, 1/2, 1$ | $0, -1/4, -1/2$ | $0, 0, 0$ | $1/2, 1/4, 0$ | $0$ | $0, 0$ | $-1$ |

Table 1: Various mechanisms as special cases of the parameterized payoff function in Equation 1. Note that the bargaining game, being asymmetric, is described by two payoff functions. These games are discussed in Section 5.

Our method considers only pure strategies. Although mixed strategies are generally required for infinite as well as finite games, there are broad classes of infinite games for which pure-strategy equilibria are known to exist. For example, Debreu (1952) shows that equilibria in pure strategies exist for infinite games of complete information with action spaces that are compact, convex subsets of a Euclidean space $\mathbb{R}^n$, and payoffs that are continuous and quasiconcave in the actions. Athey (2001) proves the existence of pure-strategy Nash equilibria for games of incomplete information satisfying a property called the single-crossing condition (SCC). These results encompass many familiar games of economic relevance, including auction games such as FPSB. Our class includes games violating SCC, for which search in the space of pure strategies may not be sufficient. Nevertheless, an ability to compute best responses for the broadest possible games is useful in itself.

The best-response algorithm takes as input a piecewise linear strategy with $K$ pieces ($K-1$ piece boundaries),

$$s(t) = \begin{cases} m_1 t + b_1 & \text{if } -\infty < t \leq c_2 \\ m_2 t + b_2 & \text{if } c_2 < t \leq c_3 \\ \ldots \\ m_{K-1} t + b_{K-1} & \text{if } c_{K-1} < t \leq c_K \\ m_K t + b_K & \text{if } c_K < t \leq +\infty, \end{cases} \quad (2)$$

represented by the vectors $\vec{c}$, $\vec{m}$, and $\vec{b}$. The piecewise-linear strategy class is sufficiently flexible to approximate any strategy, although of course the complexity of the strategy or quality of the approximation suffers as nonlinearity increases.

A two-player game is *symmetric* if both players face the same payoff function. For symmetric games we start with a single seed strategy, to be repeatedly replaced with the strategy that responds best to it.[7] Most of the examples presented in this paper are symmetric and have symmetric pure equilibria. For asymmetric games, we start with a pair of seed strategies, on every iteration computing a best response to each to get the new pair.

## 4 Existence and Computation of Piecewise Linear Best-Response Strategies

Here we present our algorithm to compute the best response to a given strategy by way of a constructive proof that in our class of games, best responses to piecewise linear strategies are themselves piecewise linear. Intuitively, the proof proceeds by first deriving an algebraic expression for expected utility against the given strategy in terms of the payoff parameters, the distribution parameters, the opponent strategy parameters, own type, and own action. By appropriate partitioning of the action space, the expected utility is expressed as a piecewise polynomial in the agent's action. We then show that the action maximizing that expression (the best response) is a piecewise linear expression of the agent's type. Finally, we establish a bound for the number of pieces in the best-response strategy.

**Theorem 1** *Given a payoff function with I regions as in Equation 1, an opponent type distribution with cdf F that is piecewise uniform with J pieces and $J-1$ piece boundaries $\{d_2, \ldots, d_J\}$, and a piecewise linear strategy function with K pieces as in Equation 2, the best-response strategy is itself a piecewise linear function with no more than $2(I-1)(J+K-2)$ piece boundaries.*

---

[7] It can be shown that symmetric games must have symmetric equilibria, although there are some symmetric games with only asymmetric *pure* equilibria (Cheng et al., 2004). Iterating from a single strategy (equivalently, a symmetric profile) does not limit the search to symmetric equilibria since a cycle of length two (given our restriction to two-player games, but regardless of whether the game is symmetric) constitutes an asymmetric equilibrium.

*Proof.* Finding the best response strategy means maximizing expected utility over the other agent's type distribution. Let $T$ be the random variable denoting the other agent's type.

First, redefine $s(t)$ to include additional redundant boundary points $\{d_2,\ldots,d_J\}$ so there are now $J+K-2$ boundary points of $s(t)$, $\{c_2,\ldots,c_{J+K-1}\}$, and

$$s(t) = \begin{cases} m_1 t + b_1 & \text{if } -\infty < t \leq c_2 \\ \ldots \\ m_{J+K-1} t + b_{J+K-1} & \text{if } c_{J+K-1} < t \leq +\infty. \end{cases}$$

We now express the expected utility, factored over the pieces of $s()$ and $u()$, as

$$EU(t,a) = E_T[u(t,a,T,s(T))] =$$
$$\sum_{i=1}^{I} \sum_{j=1}^{J+K-1} E[(\theta_i t + \rho_i a + \theta'_i T + \rho'_i (m_j T + b_j) + \phi_i \mid$$
$$c_j < T \leq c_{j+1}, \beta_i \stackrel{..}{<} a + \alpha(m_j T + b_j) \stackrel{..}{<} \beta_{i+1}]$$
$$\cdot \Pr(c_j < T \leq c_{j+1}, \beta_i \stackrel{..}{<} a + \alpha(m_j T + b_j) \stackrel{..}{<} \beta_{i+1}).$$

(We use the notation "$x_i \stackrel{..}{<} y$" to denote $x_i < y$ if $i$ is odd and $x_i \leq y$ if i is even.)

If $\alpha m_j = 0$ then the summand reduces to

$$\begin{cases} (\theta_i t + \rho_i a + (\theta'_i + \rho'_i m_j) \frac{c_j + c_{j+1}}{2} + \rho'_i b_j + \phi_i) \\ \quad \cdot (F(c_{j+1}) - F(c_j)) & \text{if } \beta_i - \alpha b_j \stackrel{..}{<} a \stackrel{..}{<} \beta_{i+1} - \alpha b_j \\ 0 & \text{otherwise.} \end{cases}$$

(The derivation of the cdf $F()$ of a piecewise uniform distribution is in the appendix of the full version.)

For the case of $\alpha m_j \neq 0$, first define $x_{ij}(a) \equiv (\beta_i - \alpha b_j - a)/(\alpha m_j)$ and $y_{ij}(a) \equiv (\beta_{i+1} - \alpha b_j - a)/(\alpha m_j)$ with $x$ and $y$ swapped if $\alpha m_j < 0$. We also introduce $mm(a,b,x) \equiv min(b, max(a,x))$.

We consider first the probability term in the summand, rewriting it as

$$p_{ij}(a) \equiv \Pr\left(\beta_i \stackrel{..}{<} a + \alpha \cdot (m_j T + b_j) \stackrel{..}{<} \beta_{i+1} \,\&\, c_j < T \leq c_{j+1}\right)$$
$$= \Pr\left(x_{ij}(a) \stackrel{..}{<} T \stackrel{..}{<} y_{ij}(a) \,\&\, c_j < T \leq c_{j+1}\right)$$
$$= F(mm(c_j, c_{j+1}, y_{ij}(a))) - F(mm(c_j, c_{j+1}, x_{ij}(a))).$$

For the expectation term in the summand, we first define

$$\overline{xy}_{ij}(a) \equiv E[T \mid mm(c_j, c_{j+1}, x_{ij}(a)) \stackrel{..}{<} T$$
$$\stackrel{..}{<} mm(c_j, c_{j+1}, y_{ij}(a))]$$
$$= \frac{mm(c_j, c_{j+1}, x_{ij}(a)) + mm(c_j, c_{j+1}, y_{ij}(a))}{2}.$$

We can now express the expected utility, $EU(t,a)$, as

$$\sum_{i=1}^{I} \sum_{j=1}^{J+K-1} (\theta_i t + \rho_i a + (\theta'_i + \rho'_i m_j)\overline{xy}_{ij}(a) + \rho'_i b_j + \phi_i) \cdot p_{ij}(a). \quad (3)$$

This expression is a piecewise second degree polynomial in $a$ and simply linear in $t$. Treating it as a function of $a$, parameterized by $t$, we can find the boundaries for the polynomial pieces (which will be expressions of $t$). This is done by setting the arguments of the maxes and mins equal and solving for $a$, yielding the following four action boundaries for each region $\{\beta_i, \beta_{i+1}\}$ in $u()$ and each region $\{c_j, c_{j+1}\}$ in $s()$:

$$c_{j+1} = y_{ij}(a) \Rightarrow a = \beta_{i+1} - \alpha \cdot (m_j c_{j+1} + b_j)$$
$$c_j = y_{ij}(a) \Rightarrow a = \beta_{i+1} - \alpha \cdot (m_j c_j + b_j)$$
$$c_{j+1} = x_{ij}(a) \Rightarrow a = \beta_i - \alpha \cdot (m_j c_{j+1} + b_j)$$
$$c_j = x_{ij}(a) \Rightarrow a = \beta_i - \alpha \cdot (m_j c_j + b_j)$$

This yields a total of at most $2(I-1)(J+K-2)$ unique action boundaries. So expected utility is now expressible as a piecewise polynomial in $a$ (parameterized by $t$) with at most $2(I-1)(J+K-2)+1$ pieces.

For arbitrary $t$, we can find the action $a$ that maximizes $EU(t,a)$ by evaluating at each of the boundaries above and wherever the derivative (of each piece) with respect to $a$ is zero. This yields up to $2(I-1)(J+K-2)+1$ critical points, all simple linear functions of $t$. Call this set of candidate actions $C$ and the corresponding set of expected utilities $EU(t,C)$. The best-response function can then be expressed, for given $t$, as $\arg\max_C(EU(t,C))$. This is a "piecewise max" of the linear functions in $C$, and so it is piecewise linear.

It remains to establish an upper bound on the resulting number of distinct ranges for $t$. We claim the size of $C$, $2(I-1)(J+K-2)+1$, is such an upper bound. To see this, first note that the piecewise max of a set of linear functions must be convex (since, inductively, the max of a line and convex function is convex). It is now sufficient to show that at most one new $t$ range can be added by taking the max of a linear function of $t$ and a piecewise linear convex function of $t$. Suppose the opposite, that the addition of one line adds two pieces. They cannot be contiguous else they would be one piece. So there must be a piece of the convex function between the two pieces of the line. This means the convex function goes below, then above, then below the line and this violates convexity. Therefore, each line in $C$ adds at most one piece to the piecewise max of $C$ and therefore the piecewise linear best response to $s()$ has at most $2(I-1)(J+K-2)+1$ pieces and thus $2(I-1)(J+K-2)$ type boundaries. $\square$

Our algorithm for finding a best response follows this constructive proof. Finding $C$ takes time $O(IJK)$. To actually find the piecewise linear function, $\arg\max_C(EU(t,C))$, we employ a brute force approach that requires $O((IJK)^2)$ time. First, we find all possible piece boundaries by taking all pairs in C, setting them equal, and solving for $t$. For each $t$ range we then compute $\arg\max_C(EU(t,C))$ and merge whenever contiguous ranges have the same argmax. As the proof shows, this will yield at most $2(I-1)(J+K-2)$ type boundaries. Thus, we have shown how to find the piecewise linear best response to a piecewise linear strategy in polynomial time. The resulting function is converted to the same strategy representation ($\vec{c}$, $\vec{m}$, $\vec{b}$) that the algorithm takes as a seed for the opponent strategy.

## 5 Examples

Here we consider existing and new games and show that our method for finding best responses can confirm or rediscover known results as well as find previously unknown equilibria.

There are many games not analyzed here to which our approach is amenable, such as the All-Pay auction (both winner and loser pay their bids; encoded in Table 1), incomplete information versions of Cournot or Bertrand games, the War of Attrition (both winner and loser pay the second highest price), and voluntary participation games (agents choose an amount to contribute for a joint good and receive utility based on the sum of both contributions; encoded in Table 1). Our approach is not needed for incentive compatible mechanisms such as the Vickrey auction, but, reassuringly, our algorithm returns the dominant strategy of truthful bidding as a best response to any other strategy in that domain.

### 5.1 First-Price Sealed-Bid Auction

We consider the first-price sealed-bid auction (FPSB) with types that are drawn from $U[0,1]$ and the following payoff function:

$$u(t,a,a') = \begin{cases} t-a & \text{if } a > a' \\ (t-a)/2 & \text{if } a = a' \\ 0 & \text{otherwise.} \end{cases}$$

In words, two agents have private valuations for a good and they submit sealed bids expressing their willingness to pay. The agent with the higher bid wins the good and pays its bid, thus receiving a surplus of its valuation minus its bid. The losing agent gets zero payoff. In the case of a tie, a winner is chosen randomly, so the expected utility is the average of the winning and losing utility.

This game can be given to our solver by setting the payoff parameters as in Table 1. The algorithm also needs a seed strategy, for which we can use the default strategy of truthful bidding (always bidding one's true valuation: $a(t) = t$ for $t \in [0,1]$). This strategy is encoded as $\vec{t} = \langle 0,1 \rangle$, $\vec{m} = \langle 0,1,0 \rangle$, and $\vec{b} = \langle 0,0,0 \rangle$ (see Section 3).

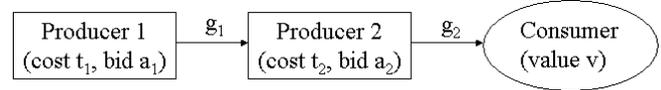

Figure 1: Supply Chain game with two producers in series.

Note that the first and last elements of $\vec{m}$ and $\vec{b}$ are irrelevant as they correspond to type ranges that occur with zero probability. After a single iteration (a fraction of a second of cpu time), our solver returns the strategy $a(t) = t/2$ for $t \in [0,1]$ which is the known Bayes-Nash equilibrium for this game (McAfee and McMillan, 1987, p709). We find that in fact we reach this fixed point in one or two iterations from a variety of seed strategies—specifically, strategies $a(t) = mt$ for $m > 0$. We approach the fixed point asymptotically (within 0.001 in ten iterations) for seed strategies $a(t) = mt + b$ with $b > 0$.

### 5.2 Supply-Chain Game

This example derives from our previous work in mechanisms for supply chain formation (Walsh et al., 2000). Consider a supply chain with two producers in series, and one consumer (see Figure 1). Producer 1 has output $g_1$ and no input. Producer 2 has input $g_1$ and output $g_2$. The consumer—which is not an agent in this model—wants good $g_2$. The producer costs, $t_1$ and $t_2$, are chosen randomly from $U[0,1]$. A producer knows its own cost with certainty, but not the other producer's cost—only the distribution (which is common knowledge). The consumer's value, $v \geq 1$, for good $g_2$ is also common knowledge.

The producers place bids $a_1$ and $a_2$. If $a_1 + a_2 \leq v$, then all agents win their bids in the auction and the surplus of producer $i$ is $a_i - t_i$. Otherwise, all agents receive zero surplus. In other words, the two producers each ask for a portion of the available surplus, $v$, and get what they ask minus their costs if the sum of their bids is less than $v$.

Walsh et al. (2000) propose a strategy for supply-chain games defined on general graphs. In the more general setting, it is the best known strategy (for lack of any other proposed strategies in the literature). For the particular instance of Figure 1, the strategy works out to:

$$a(t) = \begin{cases} t/2 + (v/2 - 1/4) & \text{if } 0 \leq t < v - 1 \\ 3t/4 + v/4 & \text{otherwise.} \end{cases}$$

Our best-response finder proves that this strategy is not a Nash equilibrium and shows how to optimally exploit agents who are playing it.

Figure 2 shows this strategy for the game with $v = (10 - \sqrt{5})/5 \approx 1.55$ (chosen so that there is a 0.9 probability of positive available surplus) along with the best response, as determined by our algorithm and confirmed by Monte Carlo simulation.

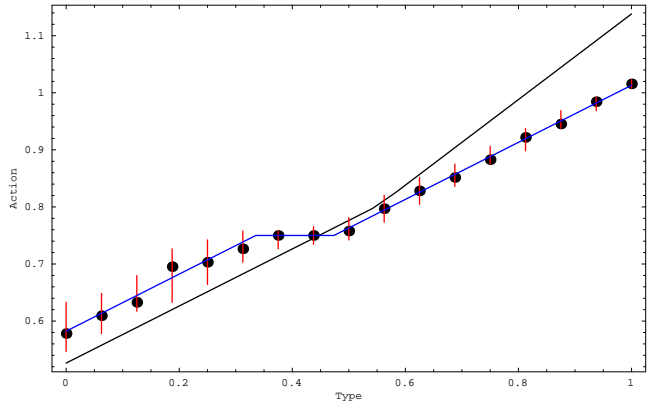

Figure 2: Hand-coded strategy for the Supply Chain game of Section 5.2, along with best response and empirical verification of best response (the error bars for the empirically estimated strategy are explained in the appendix of the full version of this paper).

When we perform further best-response iterations it eventually falls into a cycle of period two consisting of the following strategies (where $x = 3/4$):

$$a_1(t_1) = \begin{cases} x & \text{if } t_1 \leq x \\ v & \text{otherwise} \end{cases} \quad (4)$$

$$a_2(t_2) = \begin{cases} v-x & \text{if } t_2 \leq v-x \\ v & \text{otherwise.} \end{cases} \quad (5)$$

The following theorem confirms that we have found an equilibrium, and follows an analogous result (Nash, 1953) for the similar (complete information) *Nash demand game*.

**Theorem 2** *Equations 4 and 5 constitute an asymmetric Bayes-Nash equilibrium for the supply-chain game, for any $x \in [0, v]$.*

*Proof.* Assume producer 2 bids according to Equation 5. Since producer 1 cannot improve its chance of winning with a bid below $x$, and can never win with a bid above $x$, producer 1 effectively has the choice of winning with a bid of $x$ or getting nothing. Producer 1 would choose to win at $x$ precisely when $t_1 \leq x$. Hence, (4) is a best response by producer 1. By a similar argument, (5) is a best response by producer 2, if producer 1 follows Equation 4. □

Following is a more interesting equilibrium, which our solver did *not* find but we were able to derive manually and our best-response finder confirms.

**Theorem 3** *When $v \in [3/2, 3]$, the following strategy is a symmetric Bayes-Nash equilibrium for the Supply Chain game:*

$$a(t) = \begin{cases} 2/3v - 1/2 & \text{if } t \leq 2/3v - 1 \\ t/2 + v/3 & \text{otherwise.} \end{cases}$$

The appendix of the full version of this paper contains the proof which is essentially an application of our best-response algorithm to the particular game and strategy above. When this strategy is used as the seed strategy for our solver with any particular $v$, the same strategy is output, thus confirming that it is a Bayes-Nash equilibrium.

### 5.3 Bargaining Game

The supply chain game is similar to a two-player sealed-bid double auction, or bargaining game. In this game there is a buyer with value $v'$ and a seller with cost $c'$, each drawn from distributions that are common knowledge. The buyer and seller place bids and if the buyer's is greater than the seller's, they exchange the good at a price that is some linear combination of the two bids. In the supply-chain example, we can model the seller as producer 1, with $c' = t_1$. Because the consumer reports its true value, which is common knowledge, we can model the buyer as the combination of the consumer and producer 2, with $v' = v - t_2$. However, to make the double auction game isomorphic to our supply-chain example, we need to alter the game so that excess surplus is thrown away instead of shared.

The bargaining game as defined above has been well studied in the literature (Chatterjee and Samuelson, 1983; Leininger et al., 1989; Satterthwaite and Williams, 1989). We consider the special case of the bargaining game where the sale price is halfway between the buy and sell offers, and the valuations are $U[0,1]$. The payoff function for this game is encoded in Table 1.

The following is a known equilibrium (Chatterjee and Samuelson, 1983) for a seller (1) and buyer (2):

$$a_1(t_1) = 2/3t_1 + 1/4$$
$$a_2(t_2) = 2/3t_2 + 1/12$$

Our solver finds this equilibrium after several iterations (with tolerance 0.001) when seeded with truthful bidding.

### 5.4 Shared-Good Auction

Consider two agents who jointly own an inherently unsharable good and seek a mechanism to decide who should buy the other out and at what price. (For example, two roommates could use this mechanism to decide who gets the better bedroom.[8]) Assume that it is common knowledge that the agents' valuations (types) are drawn from $U[A,B]$. We propose the mechanism

$$u(t,a,a') = \begin{cases} t - a/2 & \text{if } a > a' \\ a'/2 & \text{otherwise} \end{cases}$$

which we chose because it has the property that if players bid their true valuations, the mechanism would allocate the good to the agent who valued it most and split the surplus evenly between the two (each agent would get a payoff of $t/2$ where $t$ is the larger of the two valuations). The following Bayes-Nash equilibrium also allocates the good to the agent who values it most, but that agent gets up to twice as much surplus as the other agent, depending on the minimum valuation $A$.

---
[8] Thanks to Kevin Lochner who both inspired the need for and helped define this mechanism.

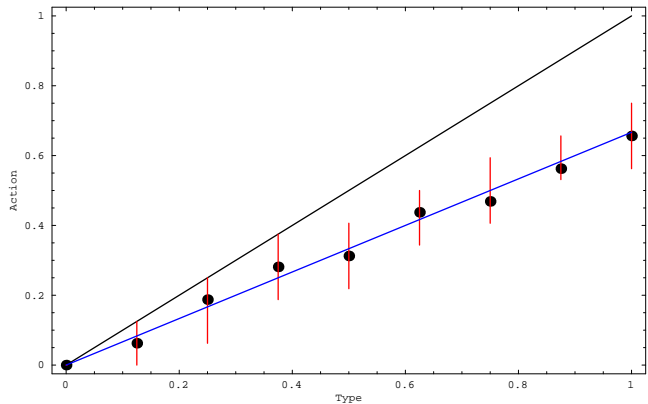

Figure 3: $a(t) = 2t/3$ is a best response to truthful bidding in the shared-good auction with $[A, B] = [0, 1]$. This strategy is in turn a best response to itself, thus confirming the equilibrium.

**Theorem 4** *The following is a Bayes-Nash equilibrium for the shared-good auction game when valuations are drawn from $U[A, B]$:*

$$a(t) = \frac{2t + A}{3}.$$

[The proof is in the appendix of the full version.]

Our solver finds this equilibrium exactly (for any specific $[A, B]$) in one iteration from truthful bidding. We confirm the result via simulation as shown in Figure 3.

### 5.5 Vicious Vickrey Auction

Brandt and Weiß (2001) introduce the following auction game:

$$u(t, a, t', a') = \begin{cases} (1-k)(t-a') & \text{if } a > a' \\ ((1-k)(t-a') - k(t'-a))/2 & \text{if } a = a' \\ -k(t'-a) & \text{otherwise.} \end{cases}$$

It is a Vickrey auction generalized by the parameter $k$ which allows agents to be "antisocial" in the sense of getting disutility from the other agent's utility. (This might be the case for businesses that are competitors.)

Brandt and Weiß derive an equilibrium only for a complete information version of this game. Our game solver can address the more general incomplete information setting.

**Theorem 5** *The following is a Bayes-Nash equilibrium for the vicious Vickrey auction game when valuations are drawn from $U[0, 1]$:*

$$a(t) = \frac{k + t}{k + 1}.$$

[The proof is in the appendix of the full version.]

Our solver finds this equilibrium (for various specific values of $k$) within several iterations from a variety of seed strategies.

## 6 Related Work

The seminal works on game theory are von Neumann and Morgenstern (1947) and Nash (1951). There are several modern general texts (Aumann and Hart, 1992; Fudenberg and Tirole, 1991; Mas-Colell et al., 1995) that analyze many of the games in Section 5. Algorithms for solving finite games include the classic Lemke-Howson algorithm (Lemke and Howson, Jr., 1964) for solving bimatrix games (two-agent finite games of complete information). In addition to the algorithms discussed in connection with GAMBIT in Section 2, there has been recent work (La Mura, 2000; Kearns et al., 2001) in algorithms for computing Nash equilibria in finite games by exploiting compact representations of games. Govindan and Wilson (2003, 2002) have recently found new algorithms for searching for equilibria in normal form and extensive form games that are faster than any algorithm implemented in GAMBIT. Blum et al. (2003) have extended and implemented these algorithms in a package called GAMETRACER. Singh et al. (2004) adapt graphical-game algorithms for the incomplete information case, including a class of games with continuous type ranges and discrete actions.

The approach of finding Nash equilibria by iterated best-response, sometimes termed *best-reply dynamics*, dates back to Cournot (1838). A similar approach known as *fictitious play* was introduced by Robinson (1951) and Brown (1951) in the early days of modern game theory. Fictitious play employs a best response, not to the single strategy from the last iteration, but a composite strategy formed by mixing the strategies encountered in previous iterations according to their historical frequency. This method generally has better convergence properties than best-response, but Shapley (1964) showed that fictitious play need not converge in general. Milgrom and Roberts (1991) cast both of these iterative methods as special cases of what they term *adaptive learning* and show that in a class of games of complete information, all adaptive learning methods converge to the unique Nash equilibrium. Fudenberg and Levine (1998) provide a good general text on iterative solution methods (i.e., learning) for finite games. Hon-Snir et al. (1998) apply this approach to a particular auction game with complete information. The *relaxation algo-* *rithm* (Uryasev and Rubinstein, 1994), applicable to infinite games, but only complete information games, is a generalization of best-response dynamics that has been shown to converge for some classes of games.

The literature is rife with examples of analytically computed equilibria for particular auction games. For example, Milgrom and Weber (1982) derive equilibria for first- and second-price auctions with affiliated signals. Gordy (1998) finds closed-form equilibria in certain common-value auctions given particular signal distributions.

## 7 Conclusion

We have presented a proof that best responses to piecewise linear strategies in a class of infinite games of incomplete information are piecewise linear. The proof is constructive and contains a polynomial-time algorithm for finding such best-responses. To our knowledge, this is the first algorithm for finding best response strategies in a broad class of infinite games of incomplete information.

For some games, this best-response algorithm can be iterated to find Bayes-Nash equilibria. It remains a goal to characterize the class of games for which iterated best-response converges. Our method confirms known equilibria from the literature (e.g., auction games such as FPSB and Vickrey), confirms an equilibrium we derive here (in the Supply Chain game), and discovers new equilibria (in the Shared Good auction and an incomplete information Vicious Vickrey auction).

**Acknowledgments**

Bill Walsh provided the original inspiration for this work—especially its application to the game in Section 5.2. The equilibrium result in that section and its proof is joint work with Bill. Terence Kelly provided many constructive suggestions and discussions. Kevin Lochner helped devise the mechanism in Section 5.4. Preliminary research for this paper was performed during an internship at Hewlett-Packard Labs and the following people at HP helped in the development of the first version our algorithm: Leslie Fine, Kay-yut Chen, Evan Kirshenbaum, Jaap Suermondt, and Kemal Guler.